\documentclass[aps,prb,twocolumn,superscriptaddress,showpacs,preprintnumbers,amsmath,amssymb]{revtex4}
\usepackage{amssymb}
\usepackage[dvips]{graphicx}
\usepackage{color}

\begin{document}

\title{Resonant escape over an oscillating barrier in single-electron ratchet transfer}

\author{Satoru~Miyamoto}
\affiliation{NTT Basic Research Laboratories, NTT Corporation, 3-1 Morinosato Wakamiya, Atsugi, Kanagawa 243-0198, Japan}
\affiliation{School of Fundamental Science and Technology, Keio University, 3-14-1 Hiyoshi, Kohoku-ku, Yokohama 223-8522, Japan}
\author{Katsuhiko~Nishiguchi}
\affiliation{NTT Basic Research Laboratories, NTT Corporation, 3-1 Morinosato Wakamiya, Atsugi, Kanagawa 243-0198, Japan}
\author{Yukinori~Ono}
\affiliation{NTT Basic Research Laboratories, NTT Corporation, 3-1 Morinosato Wakamiya, Atsugi, Kanagawa 243-0198, Japan}
\author{Kohei~M.~Itoh}
\affiliation{School of Fundamental Science and Technology, Keio University, 3-14-1 Hiyoshi, Kohoku-ku, Yokohama 223-8522, Japan}
\author{Akira~Fujiwara}
\email{afuji@will.brl.ntt.co.jp}
\affiliation{NTT Basic Research Laboratories, NTT Corporation, 3-1 Morinosato Wakamiya, Atsugi, Kanagawa 243-0198, Japan}
\date{\today}

\begin{abstract}
Single-electron escape from a metastable state over an oscillating barrier is experimentally investigated in silicon-based ratchet transfer.  When the barrier is oscillating on a time scale characteristic of the single-electron escape, synchronization occurs between the deterministic barrier modulation and the stochastic escape events.  The average escape time as a function of its oscillation frequency exhibits a minimum providing a primary signature for resonant activation of single electrons.
\end{abstract}


\maketitle

When noise-induced hopping of a Brownian particle between two stable states is subject to weak periodic perturbation, stochastic resonance takes place as a cooperative phenomenon between the noise and signal.\cite{Wiesenfeld95}  Although the concept of stochastic resonance was originally propounded as a possible explanation for ice-age periodicity, it is currently observed in a wide spectrum of nonlinear dynamic systems such as electronic circuits,\cite{Fauve83} tunnel diodes,\cite{Mantegna94} superconducting quantum interference devices (SQUIDs),\cite{Rouse94} nanoelectromechanical systems (NEMS),\cite{Badzey05} and semiconductor-based neural networks.\cite{Kasai08}  Over the past few decades, this phenomenon has received considerable attention in regard to potential applications to coherent signal amplifiers through the assistance of incoherent fluctuations that cannot be suppressed or eliminated.  It was shown that such counterintuitive behavior is due to the matching between a deterministic time scale and a stochastic one, that is, the signal period and the hopping time, respectively.\cite{Gammaitoni95,Giacomelli99}\\
\indent A large amount of theoretical work predicted an analogous phenomenon called resonant activation.\cite{Doering92,Zurcher93,Broeck93,Bier93,Brey94,Pechukas94,Hanggi94,Reimann95,Marchi95,Iwaniszewski96,Boguna98}  Particle escape from a potential well is driven when the potential barrier is oscillating on a time scale characteristic of the particle escape itself.  For an oscillation frequency much lower than the order of the escape rate, the average escape time is the mean of the crossing times over each of the higher- and lower-state barriers.  In the fast limit of the oscillation, the average escape time is the effective time required to cross the quasi-static barrier with average height.  At an intermediate frequency, the average escape time resonantly takes a minimum.  Until now, only the resonant escape of macroscopic variables has been observed in tunnel diodes\cite{Mantegna00} and current-biased Josephson junctions.\cite{Yu03}  To actualize Brownian systems on a nanoscale, the alternative use of an electron as a classical particle has been considered.\cite{Maddox92}  In particular, explored based on the motivation for current standards with a metrological accuracy, single-electron ratchet transfer devices\cite{Fujiwara08} provide us with a physical platform for investigating the nonequilibrium dynamics of single electrons in metastable states.  Recently, we suggested that an intrinsic noise arising from a thermal bath can play a significant role in single-electron ratchet transfer at 16~K.\cite{Miyamoto08}  By means of the same manner of transfer, we present here the experimental observation of single-electron resonant escape over an oscillating barrier.\\
\begin{figure}[b]
\includegraphics[width=6.865cm]{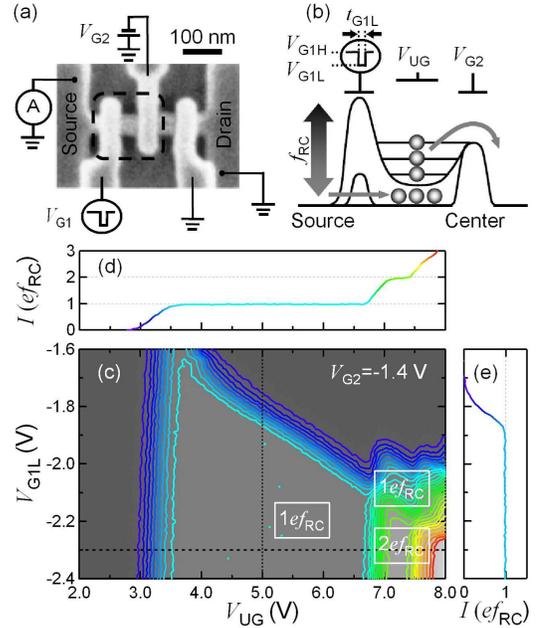}
\caption{\label{FIG1}(Color). (a) Scanning electron microscope image of the Si nanowire device mounted with triple gates before the UG formation.  (b) Schematic of the single-electron ratchet transfer employing a dynamic quantum dot enclosed by the dashed line in panel (a).  (c) Contour plots of the transfer current $I$ as a function of $V_{\textrm{UG}}$ and $V_{\textrm{G1L}}$.  (d)(e) Quantized current staircases obtained by the scans along the horizontal and vertical lines in (c).}
\end{figure}
\begin{figure}[t]
\includegraphics[width=8.285cm]{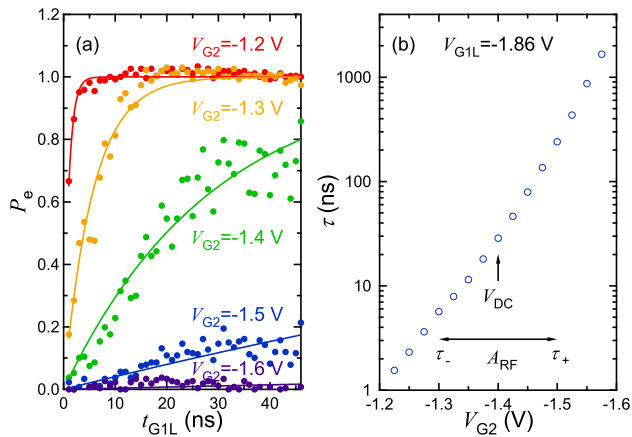}
\caption{\label{FIG2}(Color). (a) Time evolutions of the probability of electrons escaping from the metastable state $P_{\textrm{e}}$ recorded at typical values of $V_{\textrm{G2}}$.  The experimental plots are fitted with single-exponential curves (solid lines), from which the escape times $\tau$ are determined.  (b) Exponential dependence of $\tau$ on $V_{\textrm{G2}}$ obtained at $V_{\textrm{G1L}}=-1.86$~V.}
\end{figure}
\indent On a 400-nm buried oxide of a (001) silicon-on-insulator (SOI) wafer, a Si nanowire is lithographically defined with an approximate width of 30~nm and the thickness of 30~nm.  After a 20~nm-thick thermal oxide film is formed on the nanowire, it is surrounded by triple poly-Si gates.  Figure \ref{FIG1}(a) displays the top-view scanning electron microscope image of the Si nanowire metal-oxide-semiconductor field-effect transistors (MOSFETs).  Further thermal oxidation reduces the definite gate length to approximately 40~nm.  A 50~nm-thick SiO$_2$ layer is deposited on the whole device region, which is followed by the formation of a poly-Si upper gate (UG).  The wide UG layer is used as a mask during the ion implantation to form \textit{n}-type contact areas.  The application of a positive voltage to UG ($V_{\textrm{UG}}$) accumulates electrons in the undoped SOI layers left underneath UG, thereby electrically inducing source and drain on both edges of the nanowire.\\
\indent Prior to investigating the single-electron escape over the oscillating barrier, we describe the transfer scheme of the single-electron ratchet.  Now, an applied voltage to the source-side gate (G1) is pulse-modulated between $V_{\textrm{G1H}}=0$~V and $V_{\textrm{G1L}}$ at a fixed ratchet clock of $f_{\textrm{RC}}=16.67$~MHz while a constant voltage of $V_{\textrm{G2}}$ is applied to the center gate (G2).  The drain-side gate is grounded throughout this investigation.  As shown in Fig. \ref{FIG1}(b), a dynamic quantum dot formed between G1 and G2 captures single electrons from the source due to the Coulomb blockade.  The number of captured electrons $N$ can be controlled by $V_{\textrm{UG}}$.  By lifting the potential bottom sufficiently, the single electrons captured in a potential well can escape to the drain.  Repetitive transfer of single electrons produces a quantized current of $I=Nef_{\textrm{RC}}$.\cite{Fujiwara08}  Figures \ref{FIG1}(c)-(e) show the current staircases measured as a function of $V_{\textrm{UG}}$ and $V_{\textrm{G1L}}$.  The trapping of approximately one electron in each ratchet cycle can be thus brought about when $V_{\textrm{UG}}$ is set to around 5.0~V [Fig. \ref{FIG1}(d)].  However, whether or not the captured single electrons actually escape and contribute to the current depends on the height of the potential bottom controlled by $V_{\textrm{G1L}}$ [Fig. \ref{FIG1}(e)].  Then the probability that single electrons escape from the metastable state $P_{\textrm{e}}$ (less than one) can be calculated as a value of $I$ normalized by $ef_{\textrm{RC}}$.  Since approximately $10^6$ electrons are involved during the current integration, $P_{\textrm{e}}=I/ef_{\textrm{RC}}$ denotes a statistically averaged value.  The time evolution of $P_{\textrm{e}}$ can be monitored by changing the duration for which $V_{\textrm{G1L}}$ is applied, $t_{\textrm{G1L}}$.  Figure \ref{FIG2}(a) shows the time-resolved results of $P_{\textrm{e}}$ recorded at typical values of $V_{\textrm{G2}}$, which vary the height of the barrier underneath G2.  Clearly, electron escape is more likely for the lower-height barriers.  The escape time $\tau$ is determined by fitting the results with a single exponential curve.  In Fig. \ref{FIG2}(b), the obtained $\tau$ is plotted as a function of $V_{\textrm{G2}}$.  Exponential dependence of $\tau$ on $V_{\textrm{G2}}$ indicates that the escape dynamics are governed by the well-known Kramers' relation.\cite{Hanggi90}\\
\indent In order to form a dichotomously oscillating barrier as illustrated in Fig. \ref{FIG3}(a), $V_{\textrm{G2}}$ is weakly modulated at the center of $V_{\textrm{DC}}=-1.4$~V with a square-wave amplitude of $A_{\textrm{RF}}=200$~mV.  The escape rate $\tau^{-1}$ is then in the order of 1 to 100~MHz [Fig. \ref{FIG2}(b)].  When the RF frequency $f_{\textrm{RF}}$ is changed within the range from 0.16 to 158.5~MHz,\cite{Comment1} synchronization is anticipated to occur between the deterministic RF signal and the stochastic single-electron escape.
\begin{figure}[b]
\includegraphics[width=7.529cm]{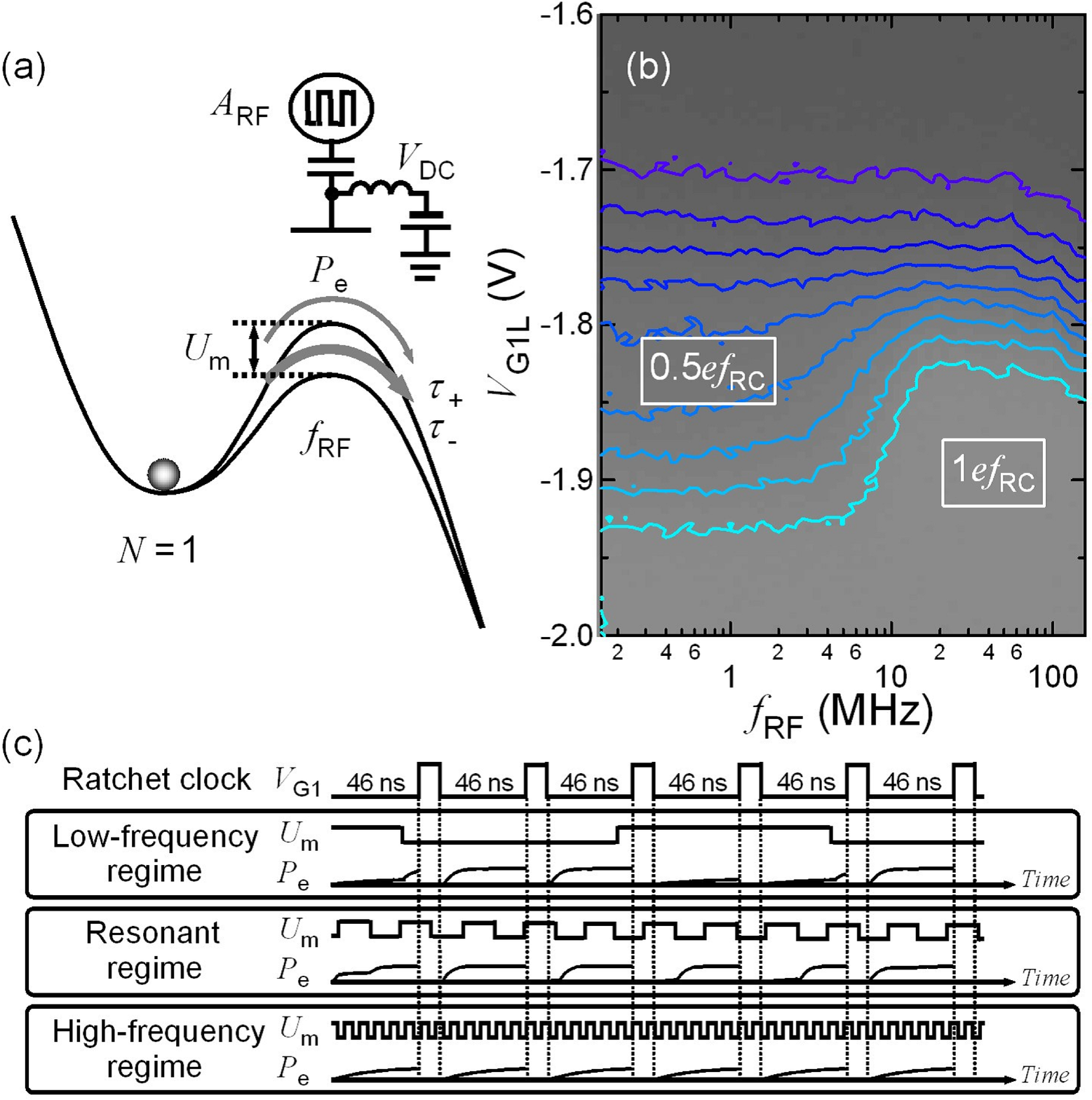}
\caption{\label{FIG3}(Color). (a) Schematic of the single-electron escape over a barrier weakly modulated by the RF signal.  (b) Current staircase along $V_{\textrm{G1L}}$ deformed by varying the RF frequency, $f_{\textrm{RF}}$.  (c) Time-sequence diagram of $P_{\textrm{e}}$ response to the dichotomous barrier modulation $U_{\textrm{m}}$ in low-frequency, resonant, and high-frequency regimes.  Out of phase with the RF signal, the single-electron ratchet transfer is independently operated in the ratchet clock period of $f_{\textrm{RC}}^{-1}=60$~ns.}
\end{figure}
\noindent Figure \ref{FIG3}(b) shows that the current staircase along $V_{\textrm{G1L}}$ is deformed by changing $f_{\textrm{RF}}$.  It is clear that the escape behavior of electrons depends on the oscillating frequency of the barrier.  A wide plateau of $P_{\textrm{e}}\sim0.5$ appears in a lower-frequency regime whereas in a higher-frequency regime the contour lines of $P_{\textrm{e}}\ge0.5$ are significantly pushed out towards a negatively smaller $V_{\textrm{G1L}}$, which indicates more efficient escape of single electrons.\\
\begin{figure}[t]
\includegraphics[width=8.500cm]{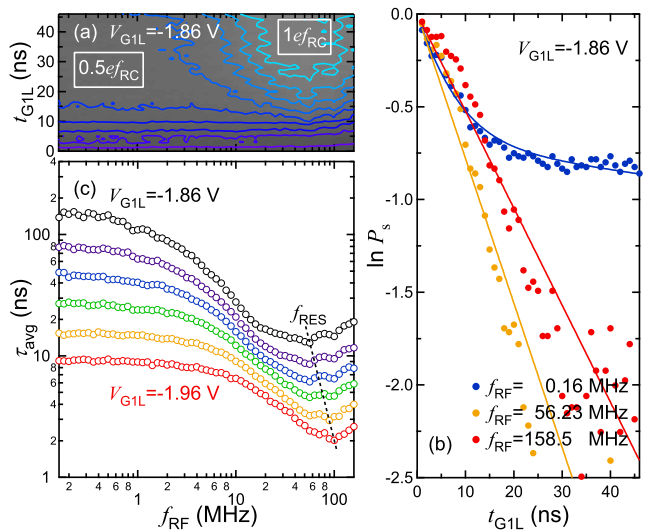}
\caption{\label{FIG4}(Color). (a) Time-resolved measurements of the escaping electrons as a function of $f_{\textrm{RF}}$.  (b) Temporally decaying probability of electrons surviving in the potential well $P_{\textrm{s}}=1-P_{\textrm{e}}$ measured at low, resonant, and high frequencies.  The solid lines are given by fitting with the single- or double-exponential functions.  (c) RF-frequency dependence of the average escape time $\tau_{\textrm{avg}}$ obtained at different values of $V_{\textrm{G1L}}$.  The dotted line indicates a correlation between the minimum values of $\tau_{\textrm{avg}}$ and the resonant frequencies $f_{\textrm{RES}}$.}
\end{figure}
\indent The escape dynamics of single electrons over the oscillating barrier are highlighted through time-resolved measurements.  Figure \ref{FIG4}(a) shows the time-domain data of escaping electrons as a function of $f_{\textrm{RF}}$.  Similar to the phenomena observed in Fig. \ref{FIG3}(b), electron escape is suppressed in the lower-frequency regime whereas it is resonantly driven by the RF signal with the $f_{\textrm{RF}}$ around several tens of megahertz.  Here, the quantity of our central interest is the average escape time defined as $\tau_{\textrm{avg}}=\int_0^\infty t_{\textrm{G1L}}\left(-dP_{\textrm{s}}/dt_{\textrm{G1L}}\right)dt_{\textrm{G1L}}$, where $P_{\textrm{s}}=1-P_{\textrm{e}}$ is the probability of electrons surviving in the potential well.  For the low-frequency regime, the escaping electrons surmount the two-height barrier slowly oscillating between the higher and lower states with a 50-50 duty cycle [Fig. \ref{FIG3}(c)].  The temporal evolution of $P_{\textrm{s}}$ shown in Fig. \ref{FIG4}(b) unambiguously exhibits the double-exponential decay at a low RF frequency of $f_{\textrm{RF}}=0.16$~MHz.  Accordingly, $P_{\textrm{s}}$ is approximately characterized by two different escape times $\tau_{+}^{\textrm{RF}}$ and $\tau_{-}^{\textrm{RF}}$ in the low-frequency regime: $P_{\textrm{s}}=\left[\exp\left(-t_{\textrm{G1L}}/\tau_{+}^{\textrm{RF}}\right)+\exp\left(-t_{\textrm{G1L}}/\tau_{-}^{\textrm{RF}}\right)\right]/2$.  Here $\tau_{\textrm{avg}}$ becomes identical to $\left(\tau_{+}^{\textrm{RF}}+\tau_{-}^{\textrm{RF}}\right)/2$.  On the other hand, when $f_{\textrm{RF}}$ is as high as the inverse of the time required for electron escape over the lower-state barrier $\tau_{-}^{-1}$, single electrons preferentially cross the lower-state barrier at least once [Fig. \ref{FIG3}(c)].  For a very high $f_{\textrm{RF}}$, single electrons experience an average-height barrier.  In these higher-frequency regimes, $P_{\textrm{s}}$ likely represents a single-exponential decay of $P_{\textrm{s}}=\exp\left(-t_{\textrm{G1L}}/\tau^{\textrm{RF}}\right)$ [Fig. \ref{FIG4}(b)].  $\tau_{\textrm{avg}}$ is then given by $\tau^{\textrm{RF}}$.  Thus, $\tau_{\textrm{avg}}$ can be obtained with good approximation by means of single- or double-exponential fitting.  In Fig. \ref{FIG4}(c), $\tau_{\textrm{avg}}$ is plotted as a function of $f_{\textrm{RF}}$.  As expected, $\tau_{\textrm{avg}}$ is found to manifest a resonance.  Such a nonmonotonic feature is robust and can be observed even when the order of $\tau_{\textrm{avg}}$ is changed by $V_{\textrm{G1L}}$.  The resonant frequency $f_{\textrm{RES}}$ exhibits a shift to higher frequencies for a shorter $\tau_{\textrm{avg}}$ and furthermore the minimum $\tau_{\textrm{avg}}$ at $f_{\textrm{RES}}$ is shifted along the dotted line.  Namely, the time-scale matching with the deterministic barrier modulation triggers the stochastic single-electron emission.  It is worth emphasizing that the observed phenomenon is clearly distinguished from the photon-assisted tunneling that can be observed when the photon energy matches or exceeds the separation between discrete levels in the potential well.\cite{Oosterkamp97}\\
\begin{figure}[t]
\includegraphics[width=8.220cm]{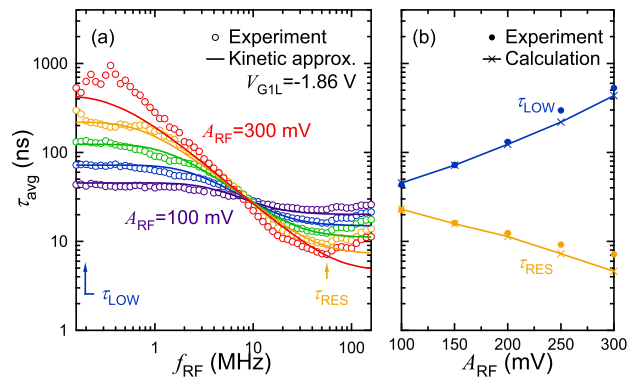}
\caption{\label{FIG5}(Color). (a) Comparison of the resonant behaviors observed as increasing the RF amplitude $A_{\textrm{RF}}$ with the kinetic approximation given by Eq. (\ref{eq1}).  $\tau_{\textrm{LOW}}$ and $\tau_{\textrm{RES}}$ are $\tau_{\textrm{avg}}$ obtained at the low-frequency limit in the present experiment and at the resonant frequency, respectively.  (b) Variations in $\tau_{\textrm{LOW}}$ and $\tau_{\textrm{RES}}$ as a function of $A_{\textrm{RF}}$.  The two values are respectively compared to $\tau_{\textrm{kin}}|_{f_{\textrm{RF}}\rightarrow0}=(\tau_{+}+\tau_{-})/2$ (upper solid line) and $\tau_{\textrm{kin}}|_{f_{\textrm{RF}}\rightarrow\infty}\approx2\tau_{-}$ (lower solid line).}
\end{figure}
\indent In addition, the resonant variation of $\tau_{\textrm{avg}}$ can be tuned by $A_{\textrm{RF}}$ as shown in Fig. \ref{FIG5}(a).  With an increase in $A_{\textrm{RF}}$, the differences in $\tau_{\textrm{avg}}$ between the lower- and higher-frequency regimes become more pronounced with the inflection points clamped around 8~MHz where $\tau_{\textrm{avg}}$ is almost the same as the value of $\tau$ at $V_{\textrm{DC}}$ [Fig. \ref{FIG2}(b)].\cite{Marchi95}  The experimental results are compared with the kinetic approximation,\cite{{Boguna98}} which is a theoretical framework effective only for a frequency regime lower than $f_{\textrm{RES}}$.  Equation (58) in Ref. 19 is given as
\begin{equation}
\tau_{\textrm{kin}}=\frac{1}{2}\left(\tau_{+}+\tau_{-}\right)-\frac{1}{2}\left(\tau_{+}-\tau_{-}\right)\frac{q_{+}-q_{-}}{1-q_{+}q_{-}},
\label{eq1}
\end{equation}
where $q_{\pm}=\exp\left(-1/2\tau_{\pm}f_{\textrm{RC}}\right)$.  In Fig. \ref{FIG5}(a) the results of the kinetic approximation are plotted, which are calculated based on Eq. (\ref{eq1}) using $\tau_{+}$ and $\tau_{-}$ estimated from Fig. \ref{FIG2}(b).  Good agreement is obtained except for a frequency regime higher than $f_{\textrm{RES}}$.\cite{Comment2}  In the high-frequency regime where single electrons surmount the average-height barrier, $\tau_{\textrm{avg}}$ should asymptotically approach an $A_{\textrm{RF}}$-independent value of $\sqrt{\tau_{+}\tau_{-}}$ equal $\tau$ at $V_{\textrm{DC}}$.\cite{Doering92}  In the present experiment, it is confirmed that $\tau_{\textrm{avg}}$ in the high-frequency regime gradually deviates from $\tau_{\textrm{kin}}$ and exhibits a slight increase.\\
\indent Finally, we discuss the physical meanings of $\tau_{\textrm{LOW}}$ and $\tau_{\textrm{RES}}$ extracted from Fig. \ref{FIG5}(a) by comparing them to the limit values of $\tau_{\textrm{kin}}$ in Fig. \ref{FIG5}(b).  $\tau_{\textrm{LOW}}$ is consistent with $\tau_{\textrm{kin}}|_{f_{\textrm{RF}}\rightarrow0}=(\tau_{+}+\tau_{-})/2$.  In this limit electron escape takes place over either the higher-state barrier or lower-state barrier exclusively with probability 1/2 for each.  Especially when $\tau_{-}<t_{\textrm{G1L}}<\tau_{+}$, electron escape is almost suppressed for the higher-state barrier, thereby giving rise to 0.5 plateaus in Figs. \ref{FIG3}(b) and \ref{FIG4}(a).   Meanwhile, $\tau_{\textrm{RES}}$ is well approximated by $\tau_{\textrm{kin}}|_{f_{\textrm{RF}}\rightarrow\infty}=2/(\tau_{+}^{-1}+\tau_{-}^{-1})\approx2\tau_{-}$.  This implies that the escaping electrons most likely cross the barrier when it is switched to the lower state.\cite{Doering92}  More specifically, even if single electrons once fail to cross the lower-state barrier, they have another chance after a half period of $f_{\textrm{RF}}^{-1}$.  Hence, the quantitative evaluation supports that the observed phenomenon can be intuitively understood as shown in Fig. \ref{FIG3}(c).\\
\indent In conclusion, the stochastic resonant escape of single electrons was experimentally verified in silicon-based single-electron ratchet transfer.  For the barrier-oscillating frequency low compared to the escape rate, the transfer current is suppressed since single electrons are inevitably subject to the higher-state barrier formed with probability 1/2.  When the barrier is oscillating at a frequency in the order of $\tau_{-}$, the majority of the escape events take place in the configuration of the lower-state barrier.  The coincidence in the characteristic time scales induces the resonant escape of single electrons, consequently enhancing the transfer current.  The origin of the observed phenomenon is the interaction of the deterministic driving signal with the stochastic behavior in Brownian systems.  The understanding of such a coordinated interaction would be of importance for noise-assisted operation of nanoelectronic devices nonisolated from a thermal bath.\\
\indent Different aspects of this work were supported by the Grant-in-Aid for Scientific Research (Grant Nos. 20241036, 19310093, and 18001002), Grant-in-Aid for the Global Center of Excellence for High-Level Global Cooperation for Leading-Edge Platform on Access Spaces from MEXT, and Special Coordination Funds for Promoting Science and Technology.


\begin{thebibliography}{}
\bibitem{Comment1}Initialization of single-electron trapping from the source is robust with respect to the continuous RF signal.
\bibitem{Comment2}The fractional deviation in the low-frequency regime is mainly due to the double-exponential fitting within the finite measurement window.
\end{thebibliography}
\end{document}